\documentclass[%
 reprint,
superscriptaddress,
 amsmath,amssymb,
 aps,
prb,
]{revtex4-1}

\usepackage{mathtools}
\usepackage{graphicx}
\usepackage{dcolumn}
\usepackage{bm}
\usepackage[colorlinks,linkcolor=red,citecolor=blue]{hyperref}

\usepackage{caption}
\usepackage{subcaption}

\begin{document}

\preprint{APS/123-QED}

\title{Collective excitations of quantum Hall states under tilted magnetic field}

\author{Kang Yang}
\affiliation{Department of Physics, Stockholm University, AlbaNova University Center, 106 91 Stockholm, Sweden}
 \affiliation{ Laboratoire de Physique Th\' eorique et Hautes Energies, CNRS UMR 7589, Sorbonne Universit\' e, 4 place Jussieu, 75252 Paris Cedex 05, France}
\affiliation{Laboratoire de Physique des Solides,  CNRS UMR
8502, Universit\' e Paris Saclay, 91405 Orsay Cedex, France}
\author{Mark Goerbig}
\affiliation{Laboratoire de Physique des Solides,  CNRS UMR
8502, Universit\' e Paris Saclay, 91405 Orsay Cedex, France}
\author{Beno\^it Dou\c cot}
\affiliation{ Laboratoire de Physique Th\' eorique et Hautes Energies, CNRS UMR 7589, Sorbonne Universit\' e, 4 place Jussieu, 75252 Paris Cedex 05, France}

\date{\today}

\begin{abstract}
We study the neutral excitations of fractional quantum Hall states in electronic systems of finite width where an external anisotropy is introduced by tilting the magnetic field. As in the isotropic case, the neutral collective excitation can be worked out through the conserving method of composite fermions in the Hamiltonian theory because the interaction potential has a natural cutoff due to the quantum well width. We show how such a computation can be carried out perturbatively for an anisotropic interaction. We find that unlike the charge gap, the neutral collective gap is much more sensitive to the tilt and can thus, for certain fractional quantum Hall states, be easily destroyed by the parallel component of the magnetic field. We also discuss the convergence of the collective spectrum to the activation gap in the large-momentum limit.
\end{abstract}

\maketitle


\section{\label{sec:level1}Introduction}

Fractional quantum Hall (FQH) states of two dimensional ($2$D) systems in a strong perpendicular magnetic field are protected by their intrinsic topological order. As topology does not rely on how distance is defined on the base manifold where the system lies in, these FQH states are expected to exhibit robustness against geometric perturbations. A possible origin of geometric perturbation is a broken rotational symmetry, making the distance in one direction longer than that in the orthogonal direction. In recent years, experimental progress \cite{xia2011evidence,PhysRevLett.108.196805,PhysRevB.88.035307,PhysRevLett.110.206801,PhysRevB.98.205418,PhysRevB.100.041112} has been achieved to induce anisotropy in these systems to probe the topologically protected FQH states, for instance by tilting the magnetic field. This becomes a powerful tool to tune the quantum Hall systems through different interesting phases.

However, most model states \cite{PhysRevLett.50.1395,PhysRevLett.63.199,MOORE1991362,PhysRevB.59.8084} for FQH systems also preserve rotational symmetry. Unlike the discrete robustness due to topology, such a continuous symmetry is  sensitive to anisotropy. In the isotropic case, a quantum Hall nematic state possessing a preferred axis can be induced spontaneously by adding pressure \cite{PhysRevB.96.035150,samkharadze2016observation}. This state breaks the rotational symmetry but maintains the same topological orders as the rotationally invariant quantum Hall liquid such as the Laughlin state and the Moore-Read state. The possibility of inducing  an anisotropic topological state under external anisotropy remains an intriguing question.

The nematic phase transition in quantum Hall systems is related to the gap closing in the collective spectrum \cite{PhysRevB.88.125137,PhysRevX.4.041050}, which is controlled by inherent strong electronic correlations. The strongly interacting problem is always difficult to solve. However, Jain's composite fermion (CF) \cite{PhysRevLett.63.199} picture allows, in the framework of the Hamiltonian theory developed by Shankar and Murthy \cite{PhysRevLett.79.4437,PhysRevLett.83.2382}, for an approximate analytical treatment of FQH states. Indeed, the FQH states of electrons at filling factors $\nu=N_e/N_\phi=p/(2ps+1)$ can be interpreted as integer quantum Hall (IQH) states of CFs at filling $\nu^\ast=p$ by attaching statistic flux to compensate partially the external magnetic field felt by CFs. These completely filled CF Landau levels (CF LLs) provide a natural reference state for a perturbative diagrammatic treatment, such as the Hartree-Fock (HF) approximation. Within the Hamiltonian theory, which is directly related to Jain's original picture, and standard flux-attachment procedure \cite{PhysRevB.44.5246} in the long-wavelength limit $ql\ll 1$, where $l=\sqrt{\hbar/eB}$ is the magnetic length, one may use the time-dependent Hartree-Fock (TDHF) approximation to study the collective excitations of CF states. In this case, the CF IQH state with $p$ filled CF LLs can be viewed as the mean-field ground state. Similarly to the electronic IQH effect \cite{PhysRevB.30.5655}, the collective excitations involve interacting CFs and CF holes, which are known as magneto-excitons. In diagrammatic language, this method considers only bubble and ladder diagrams. It conserves the gauge symmetry of the system and proves to be successful in quantum Hall systems with a good reference ground state \cite{PhysRevB.58.16262}.

In this paper, we study the collective neutral spectrum of the FQH states under anisotropy within the CF Hamiltonian theory. The anisotropy is introduced by tilting the magnetic field in a quantum well of finite width. As the Hamiltonian theory is based on a long-wavelength approximation, it provides good results in the case of a short-distance cutoff that is naturally provided in the present case by the width of the quantum well. A key feature in the presence of anisotropy is that CF LLs start to mix. We show how a perturbation theory can be constructed and use the latter to calculate the anisotropic contribution to the spectrum. The vulnerability of the collective gap, especially in CF states beyond the most robust Laughlin states, suggests a possible nematic phase transition. We also discuss the long wave-vector behaviour of the neutral spectrum obtained in the Hamiltonian theory, a point that has not been paid much attention to. We show that the TDHF formalism based on the electron density needs an infinite number of CF LLs to remain numerically valid in the large-momentum limit, while the physical constraint in the Hamiltonian theory makes the CF LLs increasingly coupled.

The structure of our paper is as follows. In Sec.~\ref{sc_hm} we review the Hamiltonian theory and especially the TDHF formalism. In Sec.~\ref{sc_cfeden} we discuss the spectra computed within two complementary approaches, one based on the physical electronic density, while the other one appeals to the so-called \textit{preferred} density, which can be interpreted as the CF density in the long-distance limit where the internal structure of the CF is discarded. We illustrate the two approaches in relation with the activation gap. Section~\ref{sc_perani} introduces  the perturbation theory and the anisotropic interaction for a tilted magnetic field. The spectrum under anisotropy is presented in Sec.~\ref{sc_anicoll}, where we compute the collective spectra along the $x-$ and the $y-$ directions for $\nu=1/3$ and $\nu=2/5$.

\section{The collective excitations in the Hamiltonian theory}\label{sc_hm}

The neutral collective excitations of FQHE states described by Laughlin's wave functions were first obtained by Girvin, MacDonald and Platzman through a single mode approximation (SMA) \cite{PhysRevLett.54.581,PhysRevB.33.2481}. In this approximation, the lowest-energy collective excitation is created by modulation of the density. The spectrum is gapped at zero momentum, in accordance with the incompressibility of FQH liquids. For the $\nu=1/3$ state, a minimum has been identified as a finite wave-vector gap, called the magneto-roton gap. This has later been confirmed by numerical studies \cite{PhysRevLett.54.237} and model wavefunction studies \cite{PhysRevLett.108.256807}, and its stability can be viewed as a tendency of the liquid to form a crystal with a period governed by the wave vector of the magneto-roton minimum.

In Jain's wave functional CF approach \cite{PhysRevLett.69.2843,PhysRevB.61.13064}, the collective excitation is constructed by superposition of trial wavefunctions with a CF in the lowest empty CF LL and a CF-hole in the topmost filled CF LL. This is reminiscent of the magneto-exciton excitations in IQH effects, worked out by Kallin and Halperin \cite{PhysRevB.30.5655}. Unlike the electron counterparts, the CF-LL structure emerges from the interaction and their spacings are of the same magnitude as the residual inter-CF two-body interaction. The CF and CF-hole are thus intrinsically strongly interacting entities. The behaviour of the spectrum is qualitatively different from the electron magneto-excitons but matches quite well with SMA results.

Here, we use an approach that is complementary to the one based on CF wavefunctions. We make use of the Hamiltonian theory by Shankar and Murthy of which the basic elements we briefly review below. Furthermore, we provide an introduction to the TDHF approximation for the time-ordered density-density correlation function, which is our principal ingredient in the determination of collective excitations.

\subsection{The Hamiltonian theory of Shankar and Murthy}

Within Jain's wavefunction approach, the CF can be viewed as composite particle that consists of an electron which is bound to a vortex of positive charge and carrying an even number of flux quanta. Shankar and Murthy formalized this picture within a second-quantized Hamiltonian theory  \cite{RevModPhys.75.1101}, in which the vortex is described as an additional degree of freedom. The inevitable redundance -- the vortices are naturally themselves composed of electrons in a strongly correlated state -- is countered by the introduction of a physical constraint to maintain the size of the original Hilbert space. To illustrate this method, we focus on the fractional fillings $\nu=p/(2ps+1)$.

Remember that the kinetic energy of 2D electrons of band mass $m$ in a strong magnetic field $B$ is given by the Hamiltonian
\begin{equation}\label{eq:ham0}
    H_0 = \frac{1}{2}m \omega_c\sum_j \left(\eta_{j,x}^2+\eta_{j,y}^2\right),
\end{equation}
in terms of the cyclotron frequency $\omega_c=eB/m$ and the components of the cyclotron variable
$(\eta_{j,x},\eta_{j,y})$ of the $j$-th electron. The non-commutativity $[\eta_{j,x},\eta_{j',y}]=il^2\delta_{j,j}$ 
is at the origin of the usual quantization into highly degenerate LLs,
$E_n=\hbar\omega_c(n+1/2)$. The LL degeneracy is related to the guiding-center coordinate, which describes the center of the cyclotron motion and thus a constant of motion that commutes with the kinetic Hamiltonian (\ref{eq:ham0}). However, the $x$- and $y-$components of the guiding center of a single electron do not commute,
\begin{equation}
    [R_{ex},R_{ey}]=-il^2,
\end{equation}
and each LL possesses a large number of states given in terms of the number of flux quanta $N_\phi=S/2\pi l^2$ threading the surface $S$. The occupation of a LL is therefore characterized by the filling factor $\nu=N_e/N_\phi$ mentioned in the introduction, where $N_e$ is the number of electrons in the 2D gas. If $\nu$ is not an integer, the LLs are only partially filled, and one is confronted with a strongly correlated electronic system. Indeed, the guiding-center coordinates of the electrons in the partially filled LL now constitute the remaining degrees of freedom, and one is left with an interaction Hamiltonian
\begin{equation}
H=\frac{1}{2}\sum_{\mathbf q}V_{\textrm{eff}}(\mathbf q)\rho_e(\mathbf q)\rho_e(-\mathbf q),\label{eqprjH}
\end{equation}
where the projected electron density operator reads $\rho_e(\mathbf q)=\sum_j\exp(-i\mathbf q\cdot\mathbf R_j)$, in terms of the guiding-center coordinate $\mathbf{R}_j$ of the $j$-th electron.
The effective interaction potential $V_{\textrm{eff}}$ is given by the combination of the bare Coulomb interaction $2\pi e^2/\epsilon q$ and the a form factor due to the projection to a single LL -- in the remainder we consider the lowest LL (LLL). Furthermore, the effective potential contains form factors that take into account the finite well width and that can become anisotropic in the presence of an inplane component of the magnetic field as we discuss in more detail in this paper. For simplicity we set $e^2/\epsilon l$ as the energy unit throughout this paper where $\epsilon$ is the dielectric constant.

The guiding center coordinates form a pair of conjugate variables, and the many-body Hilbert space of $N_e$ electrons in a LL is exponentially large. The idea of CFs is to convert the FQH problem to an IQH problem of these quasiparticles. The CF is constructed by attaching statistical flux quanta, which are opposite to the external magnetic field, to electrons and thus reducing the effective magnetic field felt by the CF. In Jain's wave unction construction, this is done by attaching vortices with $2s$ flux quanta to electrons. In the Hamiltonian theory, these vortices are described as ``independent'' particles of charge $-c^2e$, where $c^2=2ps/(2ps+1)$ and $e<0$ is the electron charge. Similarly to the electronic degrees of freedom inside a LL, the vortex dynamics is described in terms of the
guiding-center coordinates $\mathbf R_v$ the components of which  satisfy the commutation relation $[R_{vx},R_{vy}]=il^2/c^2$. These additional degrees of freedom allow us to form linear combinations of the electronic and vortex guiding-center operators,
\begin{equation}
\boldsymbol \eta=\frac{c}{1-c^2}(\mathbf R_v-\mathbf R_e) ~~\text{and} ~~ \mathbf R=\frac{\mathbf R_e-c^2\mathbf R_v}{1-c^2},\label{eqrlcfe}
\end{equation}
which can be viewed as the CF cyclotron and the CF guiding-center coordinates, respectively. They
satisfy the commutation relations $[\eta_x,\eta_y]=il^{*2}$ and $[R_x,R_y]=-il^{*2}$, while $[\eta_{x/y},R_{x/y}]=0$, in terms of an effective CF magnetic length $l^\ast=\sqrt{2ps+1}l$.
One notices that this new effective magnetic length is larger than the original one and thus corresponds to a lower effective magnetic field $B^*$ felt by the CFs. This allows one to introduce a CF filling factor $\nu^*=N_e/N_{\phi^*}=2\pi N_e l^{*2}$, which is related to the orginal electronic filling factor by
\begin{equation}
    \nu=\frac{\nu^*}{2s\nu^*+1}.
\end{equation}
Most saliently, we are now able to define a reference state that consists of $p$ completely filled CF LLs, i.e. $\nu^*=p$. Contrary to the original electronic system at a partial filling factor $\nu$, this CF state is no longer extensively degenerate and thus a good starting point for a perturbative or diagrammatic treatment that is at the heart of the Hamiltonian theory.

In spite of this appealing prospective, there is a drawback: since the vortex is a collective configuration of electrons, treating the vortex as an independent particle leads to an inevitable double counting of the degrees of freedom. The above enlarged Hilbert space thus requires the introduction of a constraint to restrict the dynamics to a subspace that corresponds to the physical degrees of freedom of the original problem. This constraint can be found to leading order in $\mathbf q$ through a renormalization flow separating electronic inter- and intra-LL physics \cite{PhysRevLett.79.4437}. It requires that the action of the vortex density operator on physical states should vanish \cite{PhysRevLett.83.2382}:
\begin{equation}
    \chi(\mathbf q)|\textrm{phys}\rangle=0,\label{eq_constr}
\end{equation}
where the pseudovortex density operator is $\chi(\mathbf q)=\sum_j\exp(-i\mathbf q\cdot\mathbf R_{vj})$.

From the above structure, one notices that the CFs have both cyclotron and guiding center degrees of freedom, similarly to the original system of electrons in a strong magnetic field. Therefore,
the CF LLs can be generated by the ladder operators, defined in terms of the components of the CF cyclotron variable,
\begin{equation}
    a=\frac{\eta^x+i\eta^y}{\sqrt 2l^\ast},\quad a=\frac{\eta^x-i\eta^y}{\sqrt 2l^\ast},
\end{equation}
with $[a,a^\dagger]=1$, while
the intra-CF LL states are spanned by the CF guiding centers, for instance the angular momentum when the symmetric gauge is employed. The effective
one-body Hilbert space of a CF is thus spanned by $|n,X\rangle$, where the first index corresponds to the CF LL and the second one corresponds to the intra-LL indices. The Fock space can be obtained immediately by the product of these one-body CF Hilbert spaces. Now the CFs fill $p$ LLs, and any excitation above this reference state requires a change in the CF-LL indice.

While approximate, the Hamiltonian theory thus provides a good interpretation of Jain's wavefunction approach. It provides a microscopic way to derive the CF picture, where the energy scale is correctly set to the Coulomb interaction instead of the kinetic energy. It is also useful in the calculation of the charge gap \cite{PhysRevB.63.085322}, the energy cost to create a well-separated quasiparticle/quasihole pair, which is equivalent to add up the energy of creating a hole in the filled CF LL and the energy of creating a CF in the empty CF LL. In that case, a preferred density $\rho_p=\rho_e-c^2\chi$ is employed in the interaction Hamiltonian Eq.~\eqref{eqprjH}. This density has the merit that its action on physical states is equal to the original electron density while gives the correct CF charge $e^\ast=(1-c^2)e$ in the long-wavelength limit. In the isotropic case, one can further show that these emergent CF LLs are indeed orthogonal, confirming the nearly free CF picture. For the neutral collective excitation, the situation is much more complicated because the distance between the CF and the CF-hole is small and they are strongly correlated. A way to analyze it is to study the density correlation function, whose poles are directly related to the collective modes. This can be achieved within the TDHF approximation, which we will discuss in the next section.

For future use in  the following sections, we provide the matrix elements of the density operators. Here, we focus on the cyclotron part of the density operator, while the formulae for the guiding centers can be worked out similarly. The matrix elements, the derivation of which can be found in Refs.~\cite{RevModPhys.75.1101,PhysRevB.98.205150}, read
\begin{equation}
\langle n_2|e^{-i\mathbf{q\cdot}\boldsymbol\eta}|n_1\rangle=\sqrt{\frac{n_2!}{n_1!}} e^{-x/2}\left(-iq_+l^\ast\right)^{n_1-n_2}L^{n_1-n_2}_{n_2}(x),\label{eq_rho_mxe}
\end{equation}
where $x=q^2l^{\ast 2}/2, q_+=(q_x+iq_y)/\sqrt{2}$ and $L^\alpha_n$ are associated Laguerre polynomials.
The result is valid for $n_1\ge n_2$. For $n_1\le n_2$, the matrix elements are obtained from the complex
conjugation $\langle n_2|e^{-i\mathbf{q\cdot}\boldsymbol\eta}|n_1\rangle=\langle n_1|e^{i\mathbf{q\cdot}\boldsymbol\eta}|n_2\rangle^*$.

\subsection{The TDHF formalism}\label{sc_TDHF}

The time-dependent Hartree-Fock approximation consists of solving the equation of motion of the correlation function by applying the Hartree-Fock approximation. As shown in Girvin, MacDonald and Platzman's computation \cite{PhysRevB.33.2481}, the $\mathbf q\to0$ spectrum is saturated by the single mode of density modulation. Generally, the spectrum of the collective excitations can be computed with the help of the time-ordered density correlation function,
\begin{equation}
    i\Pi(t,\mathbf r)=\langle\mathcal T \rho(t,\mathbf r)\rho(0,0)\rangle.
\end{equation}
Here, $\rho$ is a density operator of electrons restricted to the lowest LL. In the following section (Sec.~\ref{sc_cfeden}), we discuss two complementary representations of this density: the original electronic density, where one needs to take explicitly into account the constraint (\ref{eq_constr}), and the preferred density as an approximate scheme that allows us to get rid of the cumbersome constraint in the large-distance limit.  The equation of motion of the correlator is written as:
\begin{align}
    -i\frac{d}{dt}\langle\mathcal T \rho(t,\mathbf r)\rho(0,0)\rangle=&\langle\mathcal T [H(t),\rho(t,\mathbf r)]\rho(0,0)\rangle\nonumber\\
    &-i\delta(t)\langle\mathcal [\rho(0,\mathbf r),\rho(0,0)]\rangle.
\end{align}
The Dirac $\delta$-function stems from the time derivative with respect to the time-ordering operator $\mathcal T$. We give a time dependence to $H$ in order to remind that $H(t)$ takes the same time as $\rho(t)$ in the commutator (although $H$ itself is conserved, but this notation facilitates the ladder-operator calculations). As we are studying the dispersion of the spectrum with respect to the momentum $\mathbf q$, it is more convenient to work in the Landau gauge. The density operator takes a product form for intra- and inter-CF LL physics and can be decomposed
\begin{equation}
    \rho(\mathbf q)=\sum_{n_1,n_2}\rho_{n_1n_2}(\mathbf q) O_{n_1n_2}(\mathbf q)
\end{equation}
 into exciton operators
\begin{equation}
    O_{n_1n_2}(\mathbf q)=\sum_{X}e^{-iq_xX}d^\dagger_{n_1,X-q_yl^{\ast2}/2}d_{n_2,X+q_yl^{\ast2}/2},\label{eq_extn}
\end{equation}
where the subscript $X\pm q_yl^{\ast 2}/2$ labels the $x$-position of the state in the Landau gauge (or $y$-momentum). The expression $\rho_{n_1n_2}(\mathbf q)$ is the matrix elements of the density operator between CF LL $n_1$ and CF LL $n_2$ and the operator $O_{n_1,n_2}$ can be understood as the creation of a CF exciton with a pair momentum $\mathbf q$ that consists of a CF hole in CF-LL $n_2$ and a CF in CF-LL $n_1$. Due to the non-commutative nature of quantum Hall systems, the distance between the quasiparticle and the quasihole in $O_{n_1n_2}(\mathbf q)$ is actually proportional to their pair momentum instead of its inverse. For the electron density $\rho_{n_1n_2}(\mathbf q)$ is given by $\langle n_1|e^{-ic\mathbf{q\cdot}\boldsymbol\eta}|n_2\rangle$, which can be calculated with the help of Eq.~\eqref{eq_rho_mxe}. So in order to study the density correlation function, it is enough to study the correlation function of the CF excitons $\langle \mathcal TO_{n_1n_2}(t,\mathbf q)O_{n'_1n'_2}(0,-\mathbf q)\rangle$. For the reference state of $p$ filled  CF LLs, the equal-time commutator of the exciton operators takes an orthogonal form:
\begin{align}
    \langle[O_{n_1,n_2}(\mathbf q),O_{n'_2,n'_1}(-\mathbf q)]\rangle=&N_{\phi^\ast}\delta_{n'_1,n_1}\delta_{n'_2,n_2}\nonumber\\
    &\times [N_F(n_1)-N_F(n_2)],\label{eq_ini_mex}
\end{align}
where $N_{\phi^\ast}$ is the number of effective magnetic flux quanta felt by the CF and $N_F$ is the Fermi distribution function. For the reference state of $p$ filled CF LLs, it is given by $N_F(n)=\Theta(p-1-n)$. In the remaining part of this paper, we do not write out the explicit dependence on $q$ in excitons and simply use $O_{n_1,n_2}$.

In order to obtain the equation of motion of the correlation function, we first observe that the Hamiltonian only contains the interaction, which is a two-body operator of the form $d^\dagger_1 d^\dagger_3d_2d_4$ (plus a chemical potential term for normal ordering), where the subscripts represent symbolically the CF LL  and intra-CF LL indices $j=(n_{j},X_j)$. In order to compute the commutator between the Hamiltonian and the exciton, we need the following relation:
\begin{align}
    [d^\dagger_1 d^\dagger_3d_4d_2,d^\dagger_{1'}d_{2'}]=&
    \delta_{1'2}d^\dagger_1d^\dagger_3d_4d_{2'}-\delta_{2'1}d^\dagger_{1'}d^\dagger_3d_4d_2\nonumber\\
    &-\delta_{1'4}d^\dagger_1d^\dagger_3d_2d_{2'}+\delta_{2'3}d^\dagger_{1'}d^\dagger_1d_4d_2.\label{eq_cmrl}
\end{align}
Notice that here we do not write out the explicit dependence on $t$. All ladder operators above should be taken as $d^\dagger(t),d(t)$ as in the equation of motion. Since the two-body interaction operator appears in the form $\sum_{\mathbf q}(1/2)V_{\textrm{eff}}(\mathbf q)\rho_{12}(\mathbf q)\rho_{34}(-\mathbf q)d^\dagger_1 d^\dagger_3d_4d_2$, the Hamiltonian is invariant under the interchange $(12)\leftrightarrow (34)$. The last two terms on the right hand side of Eq. \eqref{eq_cmrl} are equal to the first two terms. We can omit them and suppress the factor $1/2$ in front of $V_{\textrm{eff}}(\mathbf q)$. The commutator is formed by monomials of two creation operators and two annihilation operators. The time-dependent Hartree-Fock approximation consists of replacing one of the terms $d^\dagger(t)d(t)$ by their expectation value $\langle d^\dagger(t)d(t)\rangle$:
\begin{align}
    d^\dagger_1d^\dagger_3d_4d_2\quad \to\quad &\langle d^\dagger_1 d_2\rangle d^\dagger_3d_4+\langle d^\dagger_3 d_4\rangle d^\dagger_1d_2-\langle d^\dagger_1 d_4\rangle d^\dagger_3d_2\nonumber\\ &-\langle d^\dagger_3 d_2\rangle d^\dagger_1d_4.\label{eq_hf_repl}
\end{align}
If there is no CF LL mixing at the HF level, the occupation is conserved $\langle d^\dagger(t)d(t)\rangle=\langle d^\dagger(0)d(0)\rangle$ and proportional to a Kronecker $\delta$-function times the Fermi distribution function. Then we have the following result:
\begin{align}
     [d^\dagger_1d^\dagger_3d_4d_2,d^\dagger_{1'}d_{2'}]\quad \to &\quad N_F(1')\delta_{2'1}(\delta_{1'4}d^\dagger_3d_2-\delta_{1'2}d^\dagger_3d_4)\nonumber \\
    &+N_F(2')\delta_{1'2}(\delta_{12'}d^\dagger_3d_4-\delta_{32'}d^\dagger_1d_4)\nonumber \\
    &+N_F(2)\delta_{2'1}\delta_{32}d^\dagger_{1'}d_4\nonumber \\
    &-N_F(1)\delta_{1'2}\delta_{14}d^\dagger_3d_{2'}.\label{eq_av_TDHF}
\end{align}

A useful quantity to simplify the notations is the CF-LL energy $\epsilon_n$, defined as the energy to create a CF in an empty CF LL or the negative energy to create a CF-hole in a filled CF LL,
\begin{equation}
    \epsilon_n=\begin{cases}\langle\mathbf p|d_nH d^\dagger_n|\mathbf p\rangle-\langle\mathbf p|H |\mathbf p\rangle,& n>p;\\
    -\langle\mathbf p|d^\dagger_nH d_n|\mathbf p\rangle-\langle\mathbf p|H |\mathbf p\rangle, & n\le p.
    \end{cases}
\end{equation}
This energy scale also appears if we work out the HF Hamiltonian of the system, as we  do in Sec.~\ref{sc_HFH} . It describes the properties of a single quasiparticle/hole excitation as an emergent free-particle structure, similarly to Fermi liquid theory. This is a consequence of the reference state of $p$ filled LLs that one obtains directly within the Hamiltonian theory, as discussed in the previous section.

Combining all of our calculation and transforming from time to frequency space, the equation of motion of the exciton correlation function reads
\begin{align}
    \omega \langle O_{n_1n_2}O_{n^{\prime\prime}_2n^{\prime\prime}_1}\rangle=&\left(\epsilon_{n_1}-\epsilon_{n_2}\right)\langle O_{n_1n_2}O_{n^{\prime\prime}_2n^{\prime\prime}_1}\rangle+[N_F(n_2)\nonumber\\
    &-N_F(n_1)]\sum_{n'_1,n'_2}\left(V^{(2)}_{n_2,n'_1,n'_2,n_1}\right.\nonumber\\
    &-\left.V^{(1)}_{n_2,n'_1,n_1,n'_2}\right)\langle O_{n'_1n'_2}O_{n^{\prime\prime}_2n^{\prime\prime}_1}\rangle-iN_{\phi^\ast}\nonumber\\
    &\times\delta_{n_1,n^{\prime\prime}_1}\delta_{n_2,n^{\prime\prime}_2}[N_F(n_1)-N_F(n_2)].\label{eq_ve_iso}
\end{align}
where for future convenience we introduce the notations $V^{(1)}$ and $V^{({2)}}$ for the exchange and direct interactions, respectively,
\begin{align}
    V^{(1)}_{n_1,n_2,n_3,n_4}(q_x,q_y)=&\frac{1}{2\pi}\int \frac{dk_xdk_y}{2\pi}V_{\textrm{eff}}(\mathbf k)e^{i\mathbf k\times \mathbf ql^{\ast2}}\nonumber\\
    &\times \rho_{n_2n_3}(\mathbf k)\rho_{n_1n_4}(-\mathbf k),\label{eq_ve_vexc1}\\
    V^{(2)}_{n_1,n_2,n_3,n_4}(q_x,q_y)=&\frac{1}{2\pi l^{\ast2}}V_{\textrm{eff}}(\mathbf q)\rho_{n_2n_3}(\mathbf q)\rho_{n_1n_4}(-\mathbf q).\label{eq_ve_vdir1}
\end{align}
Equation~\eqref{eq_ve_iso} is an inhomogeneous matrix equation with the equal-time commutator as the source. A vector is labelled by the double indices $(n_1n_2)$.

To finish this section, we discuss the relevance of different exciton operators $O_{n_1,n_2}$. According to Eq.~\eqref{eq_ini_mex}, they can be classified into three types, depending on how they act on the reference state of $p$ filled LLs:
\begin{enumerate}
     \item $n_1\ge p,n_2<p$ or $n_1< p,n_2\ge p$, these $O_{n_1,n_2}$ create excitations across the Fermi-level, the $p$-th CF LL. Their action on the reference state of $p$ filled CF LL $|\mathbf p\rangle,\langle \mathbf p|$ is non-zero.\label{tp_phy}
     \item $n_1\ge p,n_2\ge p$ and $n_1<p, n_2<p$, these operators annihilate $|\mathbf p\rangle$ and $\langle \mathbf p|$. If the reference state is the true ground state, those operators do not enter the density correlation function.\label{tp_un}
      \item $n_1=n_2$, these $O_{n_1,n_2}$ create and annihilate the CFs in the same LL. Its average energy is zero. Their action on the reference state is also vanishing for $\mathbf q\ne 0$.\label{tp_uns}
\end{enumerate}

\section{From composite fermions to electrons}\label{sc_cfeden}

In the Hamiltonian theory, two kinds of density operators have been involved. One is the electron density $\rho_e$, which has the charge of an electron and derives from the original Hamiltonian. Another one is the preferred density $\rho_p=\rho_e-c^2\chi$, which acts on the physical states in the same manner as the electron density due to the constraint \eqref{eq_constr}. It carries the effective charge of a CF and is thus the appropriate operator to compute the activation gap of a well separated quasiparticle/hole pair \cite{RevModPhys.75.1101}. In the formalism using the preferred density, the constraint on the Hilbert space is neglected, which provides a picture that the CFs behave like free particles at long distances, where CFs essentially interact via a Coulomb potential in terms of the effective CF charge $e^\ast=(1-c^2)e$. One may wonder what predictions the two densities give on the collective mode. The preferred density to leading order in $q$ was used by Murthy \cite{PhysRevB.60.13702} to obtain the magneto-exciton spectrum, which turns out to be very different from the dispersion obtained in the SMA, whereas the results using electron density \cite{PhysRevB.64.195310} agrees much better with the SMA dispersion. On the other hand the preferred density gives the correct estimation of the activation gap while the electron density does not do so in a simple HF calculation (as we show below, see also Ref.~\cite{PhysRevB.64.195310}). It is interesting to see whether the magneto-exciton dispersion obtained by the electron density can be consistently extended to the activation gap obtained through the preferred density. In this section, we compare the two approaches and examine some general properties when the two densities are used.

\subsection{Zero mode in the electron TDHF formalism}

When the electron density is employed, it is necessary to incorporate the constraint $ \chi(\mathbf q)|\textrm{phys}\rangle=0$ since the electron is in strongly interacting. This is reflected in Murthy's result that showed that only when the constraint is taken into account one obtains a collective spectrum consistent with the SMA or CF wavefunctions \cite{PhysRevB.60.13702,PhysRevB.64.195310}. In the TDHF formalism, the constraint behaves as a zero mode and does not contribute to the density correlation function, as shown below. Moreover, this zero mode serves as a practical criterion whether the numerical solution of the equation of motion is valid.

By construction, we have $[\rho_e(\mathbf q),\chi(\mathbf q')]=0$, which implies that the Hamiltonian commutes with the pseudovortex density $[H,\chi]=0$. From these commutation relations, we find that the correlation function between the electron density and the pseudovortex density vanishes:
\begin{equation}
    \langle\mathcal T \chi(t,\mathbf q)\rho_e(0,\mathbf q')\rangle=0.
\end{equation}
So there is always a zero mode solution to the exciton correlation function:
\begin{equation}
    -i\frac{d}{dt}\sum_{n_1,n_2}\chi_{n_1n_2}\langle \mathcal TO_{n_1,n_2}(t)O_{n'_1,n'_2}(0)\rangle=0,\label{eq_csm_phyc}
\end{equation}
where $\chi_{n_1n_2}=\langle n_1|\exp(-i\mathbf q\cdot\mathbf R_v)|n_2\rangle$ is the matrix element of the pseudovortex operator between CF LL $n_1$ and CF LL $n_2$. This is evidence for the pseudovortex operator to ``vanish'' as required by the physical constraint.

Now we show that this unphysical zero mode does not enter the density correlation function and the physical constraint is reimposed self-consistently. According to our previous classification below Eq.~\eqref{eq_ini_mex}, type-\ref{tp_uns} and type-\ref{tp_un} excitons act differently on the reference state from type-\ref{tp_phy} excitons. We group all our matrices and vectors into a block form of two, one for type-\ref{tp_phy} excitons and another one for type-\ref{tp_uns} and type-\ref{tp_un} excitons. Eq.~\eqref{eq_ve_iso} takes the block form:
\begin{equation}
     \left(\begin{array}{ccc}
        \omega-A & -B  \\
        0 & \omega-D
    \end{array}\right) \left(\begin{array}{ccc}
        a & b  \\
        c & d
    \end{array}\right)=
     \left(\begin{array}{ccc}
        \tilde I & 0 \\
        0 & 0
    \end{array}\right),
\end{equation}
where $\tilde I$ is the source matrix for the equation with elements $-iN_{\phi^\ast}\delta_{n_1,n^{\prime\prime}_1}\delta_{n_2,n^{\prime\prime}_2}[N_F(n_1)-N_F(n_2)]$. The symbol $\omega$ in the matrices is an abbreviation for the frequency times the identity matrix $\omega I$. By performing the block multiplication directly, we can see that the block $c$ has to vanish as long as $\omega$ is not an eigenvalue of $D$. The elements $b$ and $d$ vanish identically because the type-\ref{tp_uns} and type-\ref{tp_un} excitons annihilate the state of $p$ filled CF LLs. Therefore, the exciton correlation function is only non-zero for those belonging to $a$ and $a=(\omega-A)^{-1}\tilde I$. Furthermore, the poles of the exciton correlation function are the eigenvalues of the matrix $A$, i.e. of the matrix
\begin{align}
    A_{n_1n_2,n'_1n'_2}=&\left(\epsilon_{n_1}-\epsilon_{n_2}\right)\delta_{n_1n'_1}\delta_{n'_2n_2}+[N_F(n_2)\nonumber\\
    &-N_F(n_1)]\left[\frac{V_{\textrm{eff}}(\mathbf q)}{2\pi l^{\ast2}}\rho_{e,n'_1n'_2}(\mathbf q)\rho_{e,n_2n_1}(-\mathbf q)\right.\nonumber\\
    &\left.-\int\frac{d^2\mathbf k}{(2\pi)^2}V_{\textrm{eff}}(\mathbf k)\rho_{e,n'_1n_1}(\mathbf k)\rho_{e,n_2n'_2}(-\mathbf k)\right.\nonumber\\
    &\times \left.e^{i\mathbf k\times\mathbf q l^{\ast2}}\right],\label{eq_mx_mexc}
\end{align}
with the double indices $(n_1,n_2)$ and $(n'_1,n'_2)$ restricted to the subspace of type-\ref{tp_phy} excitons where $[N_F(n_2)-N_F(n_1)]\ne 0$ and $[N_F(n'_2)-N_F(n'_1)]\ne 0$. The density correlation function is a linear combination of the exciton correlation function. Let us assume that the matrix $\omega-A$ is diagonalized by a matrix $P$ made of its right eigenvectors:
\begin{equation}
    \omega-A=P\Sigma(\omega) P^{-1},
\end{equation}
where $\Sigma(\omega)$ is diagonal with eigenvalues of $\omega-A$ as its diagonal entries. $P^{-1}$ is comprised by the rows of $A$'s left eigenstates. The density correlation function can be written as:
\begin{equation}
    \langle\mathcal T \rho(\mathbf q)\rho(-\mathbf q)\rangle=
    \rho^{(n)} P\Sigma(\omega)^{-1}P^{-1}\tilde I\rho^{(n)\ast},\label{eq_dc_iso}
\end{equation}
where $\rho^{(n)}=\rho_{e,n_1n_2}(\mathbf q)$ is the vector formed by the matrix elements of the electron density. Therefore the poles of the density correlation function are given by the eigenvalues of the matrix $A$, Eq. \eqref{eq_mx_mexc}.

One can then verify by straightforward calculation that $\chi_{n_1n_2}$ is indeed a left zero-eigenstate of the matrix Eq. \eqref{eq_mx_mexc}. Moreover $\chi_{n_1n_2}$ is also an eigenstate of the matrix $A$, the subspace of type-\ref{tp_phy} excitons. This is seen by observing
\begin{equation}
    (\chi_1,\chi_2)\left(\begin{array}{ccc}
        A & B  \\
        0 & D
    \end{array}\right)=0 \Rightarrow \chi_1A=0,
\end{equation}
where the vector $\chi_{n_1n_2}$ is also decomposed into the block $(\chi_1,\chi_2)$ according to our classification of excitons. Now we illustrate how this mode decouples from Eq. \eqref{eq_dc_iso}. With $\chi_1$ being a zero eigenvector of $A$, noticing that $P^{-1}$ is made of the left-row eigenvectors of $A$, the vector $\chi_1$ appears in it. Due to the commutation $\langle[\rho_e(\mathbf q),\chi(\mathbf q')]\rangle=0$, we see that the zero mode is orthogonal to the density matrix via $\chi_1\tilde I\rho^{(n)\ast}=0$. In Eq. \eqref{eq_dc_iso}, the unphysical mode $\chi_1$ in $P^{-1}$ thus does not enter this decomposition, which indicates that the pseudovortex mode does not interfere with the physical ones.

\subsection{Large wave-vector limit of the neutral excitation}

In this part we focus on the large wave-vector limit of the neutral excitation. A quite relevant issue is the convergence to the activation gap. In the Hamiltonian theory, the activation gap is computed with the help of the preferred density instead of the electron density and is quantitatively comparable to that obtained from numerical calculations within Jain's CF wavefunction approach, especially in quantum wells with a non-zero thickness. The gap is equal to the CF-LL spacing computed from the preferred density whereas the CF-LL spacing calculated with the electron density does not give the correct answer. Here we look at how this may be reconciled with the collective neutral spectrum calculation.

The activation gap is the energy to create a pair of well-separated quasiparticle and quasihole. In quantum Hall systems, since the momentum in the $x$-direction is proportional to the distance in the $y$-direction (as it is apparent in the Landau gauge), the distance between the quasiparticle and the quasihole is proportional to the pair momentum $q$ [also reflected in the form of the exciton operator Eq.~\eqref{eq_extn}]. Their energy $E(\mathbf q)$ should thus converge to the activation gap $\Delta$ in the $\mathbf q\to \infty$ limit, where the CF and the CF hole can essentially be viewed as free charge-carrying quasiparticles that contribute to the electrical transport. Below we show that in the Hamiltonian theory, the calculations based on the electron density and preferred density exhibit entirely different behaviours in the $\mathbf q\to \infty$ limit. The former needs an infinite number of CF LLs to maintain its validity while the latter converges rapidly to the activation gap. The reason comes from the necessity of the physical constraint for $\rho_e$, which can be regarded as a gauge symmetry. This gauge symmetry imposes correlation between electrons, whose nature $\cite{PhysRevLett.79.4437}$ can be traced back to the CF transformation where the Laughlin Jastrow factor is separated from the CF wave function.

The basic elements in the conserving method is the exciton operator $O_{n_1,n_2}(\mathbf q)$, creating a hole in LL $n_2$ and a CF in LL $n_1$. A relevant energy associated with the exciton is the spacing between CF LL $n_2$ and CF LL $n_1$:
\begin{equation}
    \epsilon_{n_1}-\epsilon_{n_2},
\end{equation}
defined as the energy difference between two CF LLs. The above term appears in the TDHF equation [Eq.~\eqref{eq_ve_iso}] as the diagonal matrix element. Different exciton operators are coupled through the direct and exchange interactions $V^{(2)}$ and $V^{(1)}$. Since in practical computations, a truncation on the number of CF LL is necessary to make the equation solvable numerically, we now  focus on the asymptotic behaviour of $V^{(2)}$ and $V^{(1)}$ for large momentum limited to CF LLs $n\le n_{max}$, with some $n_{max}$ as the truncation.

The direct interaction $V^{(2)}(\mathbf q)$ is easily seen to decay exponentially when $\mathbf q\to\infty$. It does not couple different excitons at large momentum. Let's now focus on the behaviour of $V^{(1)}(\mathbf q)$. For the isotropic Coulomb potential, after performing the angular integral, the integrand is a combination of a Bessel function and the decaying Gaussian factor:
\begin{align}
    V^{(1)}_{n_1,n_2,n_3,n_4}(\mathbf q)=&\int d\theta dk\sum_{m,n} c_{m,n}(n_1,n_2,n_3,n_4) e^{in\theta} \nonumber\\
    &\times(kl^\ast)^{m}e^{-iqkl^{\ast 2}\sin(\theta-\bar\theta)}e^{-\frac{k^2l^{\ast 2}}{2}}\nonumber\\
    =&\int^\infty_0 dk\sum_{m,n}c_{m,n}(n_1,n_2,n_3,n_4)2\pi \nonumber\\
    &\times J_n(qkl^{\ast 2})e^{-\frac{k^2l^{\ast 2}}{2}}(kl^\ast)^me^{in\bar\theta}\label{eq_in_vexc}
\end{align}
where $J_n$ is the Bessel function, $q=|\mathbf q|, \bar\theta=\textrm{arg }\mathbf q$ and $c_{m,n}(n_1,n_2,n_3,n_4)$ are some constants depending on the subscripts of the exchange interaction. In the above expansion $m$ and $n$ must have the same parity, because $k\to-k$ is equivalent to $\theta\to\theta+\pi$. Such an integral can be worked out exactly as a confluent hypergeometric function:
\begin{align}
    V^{(1)}_{n_1,n_2,n_3,n_4}(\mathbf q)=&\sum_{m,n}\frac{1}{l^\ast}c_{m,n}(n_1,n_2,n_3,n_4) e^{in\bar\theta}2^{(1+m-n)/2}  \nonumber\\
    &\times \pi\frac{\Gamma[(1+m+n)/2)]}{\Gamma(1+n)}(ql^\ast)^n\nonumber\\
    &\times\prescript{}{1}{F}_1\left[\frac{1}{2}(1+m+n),1+n,-\frac{q^2l^{\ast 2}}{2}\right].\label{eq_dir_hyp}
\end{align}

The leading term in the series decays $1/q$ for large $\mathbf q\to\infty$. So if the series Eq. \eqref{eq_dir_hyp} converges sufficiently fast, the exchange interaction $V^{(1)}$ also goes to zero as the separation of the quasiparticle/quasihole pair becomes larger and larger. It is nevertheless dominant over $V^{(2)}$ in this limit. As a consequence, if one fixes a finite truncation $n_{max}$ on the number of CF LLs, the different exciton operators in Eq.~\eqref{eq_ve_iso} become decoupled for large $\mathbf q$. The eigenvalues of the vertex equation Eq. \eqref{eq_mx_mexc} converge to the energy difference of two CF LLs, $\epsilon_{n_1}-\epsilon_{n_2}$.

The numerical results based on the preferred density fits very well with the above picture. The calculation becomes stable very quickly when increasing the number $n_{max}$ and the spectrum converges to the activation gap for $q\to \infty$. The $1/q$ behaviour of $V^{(1)}(\mathbf q)$ prevails the inclusion of more and more CF LLs, i.e. $\sum_{n'_1n'_2}V^{(1)}_{n_2,n'_1,n_1,n'_2}O_{n'_1n'_2}$ converges quickly. However, for the calculation based on the electron density, this is not the case because one must keep in mind the presence of the unphysical zero mode. The pseudovortex density operator $\chi(\mathbf q)$ can be treated as the generator of a gauge symmetry. All physical quantities should commute with this generator. But the weight of this generator is shifted to higher CF LLs when $q$ increases. This can be seen from the structure of this zero mode:
\begin{align}
    \chi_{n_1n_2}(\mathbf q)=&\sqrt{\frac{n_2!}{n_1!}} e^{-q^2l^{\ast2}/(2c^2)}\left(-i\frac{q_+l^\ast}{c}\right)^{n_1-n_2}\nonumber\\
    &\times L^{n_1-n_2}_{n_2}\left(\frac{q^2l^{\ast2}}{2c^2}\right).\label{eq_psmx2}
\end{align}
When $q$ is large, each component $\chi_{n_1,n_2}$ is controlled by a power term $q^{n_1+n_2}$ times an exponentially decaying factor $\exp[-q^2l^{\ast2}/(2c^2)]$. As a result, the higher $n_1+n_2$ is, the larger the component $\chi_{n_1n_2}$ is. The most significant contribution comes from $n>q/\ln{q}$ CF LLs. This pseudovortex operator, viewed as a gauge symmetry generator, thus couples lower CF LLs to very high CF LLs. In order to keep all physical quantities invariant under this gauge symmetry in a numerical computation, we have to keep around $n_{max}\sim q/\ln{q}$ CF LLs at the same time as we increase $q$. This makes the above convergence arguments fail because the dimension of the matrix $A$ goes to infinity as well. So the collective spectrum $E(\mathbf q)$ computed from $\rho_e$ does not necessarily converge to the corresponding CF-LL spacing. It can only be worked out numerically by considering more and more CF LLs for larger $q$.

\begin{figure}
    \centering
    \includegraphics[width=0.95\linewidth]{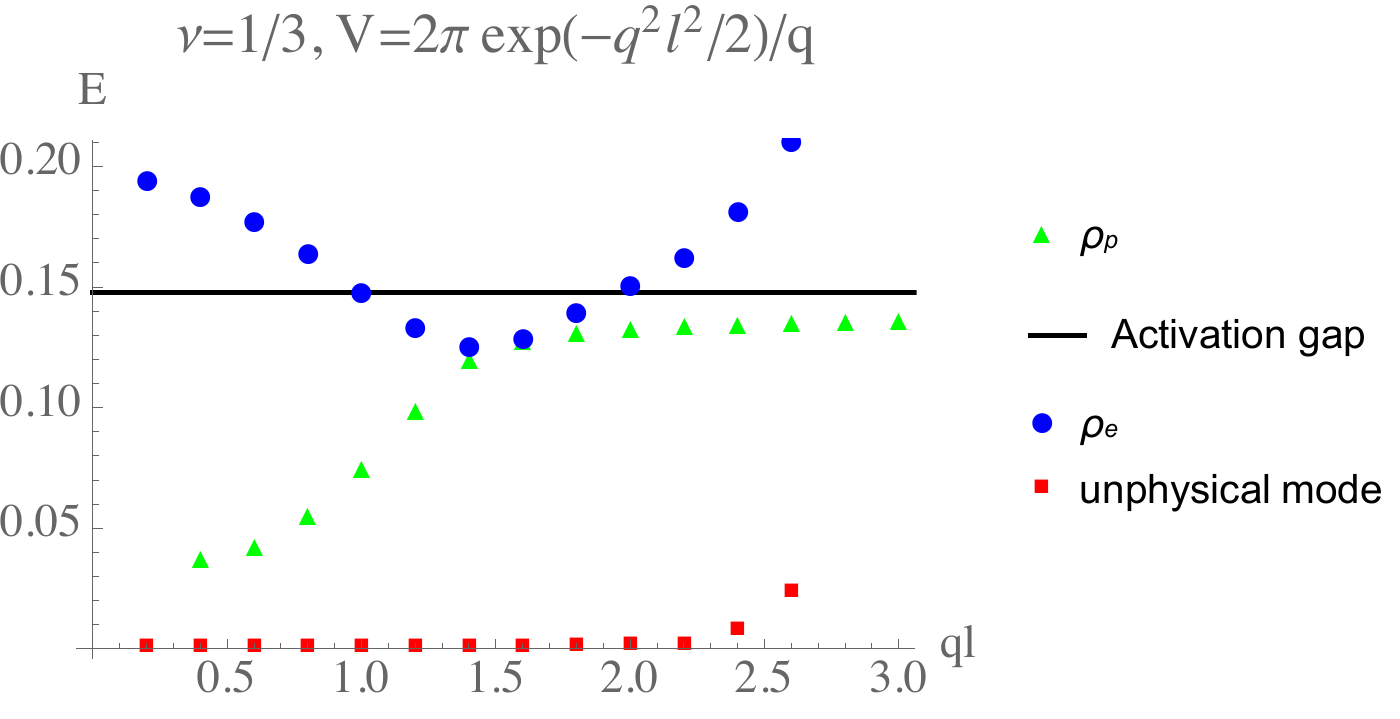}
    \caption{Comparison between the dispersion computed from the electron density (blue circles) that obtained from the preferred density (green triangles). The number of CF LLs used for $\rho_e$ is $20$, which guaranties the unphysical mode (red squares) to remain at zero energy up to $ql\sim 2.2$, while for $\rho_p$ $7$ LLs is used. The activation gap (black line) is computed using $\rho_p$. }
    \label{fig_mg_cp}
\end{figure}

Such a behaviour is also reflected in our calculation. From Fig.~\ref{fig_mg_cp}, we observe that the collective excitation based on $\rho_p$ converges rapidly to the activation gap in the large-$q$ limit. However, its small-momentum behaviour is incorrect when compared to well-established results obtained within the SMA or the CF wavefunctions, while for the calculation based on $\rho_e$, its small-momentum indeed reproduces well the SMA shape in this limit. The latter approach based on the electronic density $\rho_e$ therefore seems better adapted in the description of the collective-excitations spectrum but has a relatively slow convergence in the large-$q$ limit, where we need to keep an increasing  number of CF LLs. Otherwise the energy of the collective excitations diverges and this divergence is accompanied by a deviation from zero energy of the unphysical mode. For $q\sim 1/l$, $7$ CF LLs are sufficient to make the energy stable, but for $q\sim 2/l$, $20$ CF LLs are necessary. The unphysical zero mode therefore serves as the benchmark for the validity of the finite CF-LL truncation.

\section{The anisotropic interaction for tilted magnetic field}\label{sc_perani}

In the case of an anisotropic interaction, an important feature is that the CF LLs are mixed on the HF level \cite{PhysRevB.98.205150}. In a standard free-fermion system, level mixing can be eliminated with the help of a basis transformation in the Hilbert space. But in the case of CF LLs, the free-particle structure is only an emergent one from interactions. One cannot get rid of the mixing cannot by choosing another one-body basis. An economical method to deal with such mixing term is to use perturbation theory, which is the object of this section.

In a first step, we need to define what the perturbative parameter and show how the perturbation theory is established. We consider a generic anisotropic interaction with inversion symmetry ($\mathbf q\to -\mathbf q$) and a reflection plane ($q_x\to -q_x$). According to Fourier series expansion, the angular part of the interaction is a periodic function of the angle $\phi=\arctan(q_y/q_x)$ and thus can be expanded as
\begin{equation}
    V(\mathbf q)=\sum_n V_n(q)\cos(2n\phi),\label{eq_in_expan}
\end{equation}
where $V(q)$ is a function only of the norm of $\mathbf q$ and $\phi$ is the angle of the polar coordinate in $xy$-plane. When the interaction is isotropic, only the $V_0$ component is non-zero. This expansion is a simplified version of the generalized pseudopotential expansion \cite{PhysRevLett.118.146403}, where the radial part is also expanded according to associated Laguerre polynomials. For these generalized pseudopotentials, the radial expansion corresponds to different electronic angular-momentum states, but here, the angular expansion is enough because we study CFs that have a different magnetic length from electrons and we are mainly interested in their inter-LL physics. Now we turn on the anisotropy. From the perturbation point of view, we start from an interaction with only the $V_0$ component and then switch on small $V_{n\ne 0}$ components. As we show below, these $V_{n\ne 0}$ components perturb the CF picture since they introduce CF LL mixing. This naturally leads us to treating the CF LL mixing as a perturbation. The amplitudes of the components $V_{n\ne 0}$ serve as the perturbation parameter.

In order to illustrate the emergence of an anisotropic interaction, let us consider the quantum Hall system in a quantum well with finite thickness in the perpendicular $z$-direction under a tilted magnetic field, i.e. a non-zero $B$-field component in the $xy$-plane. We choose the inplane component in the $x$-direction and $\mathbf B=B_x\mathbf e_x+B_z\mathbf e_z$ with $B_x=B_z\tan\theta$. A simple solvable model consists of a parabolic confining potential in the $z$-direction
\begin{equation}
H_z=\frac{\Pi_z^2}{2m}+\frac{m\Omega^2 z^2}{2}.
\end{equation}
with $\Omega$ characterizing the confining strength. By tilting the magnetic field, the cyclotron motion of electrons is also tilted away from the $xy$-plane, leading to an anisotropic repulsion. The effective interaction is found to be \cite{PhysRevB.87.245315,PhysRevB.98.205150}
\begin{align}
V_{\textrm{eff}}(\mathbf q)=&\int \frac{dq_z}{2\pi}\frac{4\pi e^2}{q^2}e^{-\frac{1}{2}\left[\frac{ q_y^2\sin^2\tilde\theta}{l_+^2/l^4}+l^2_+(q_x\sin\tilde\theta-q_z\cos\tilde\theta)^2\right]}\nonumber\\
&\times e^{-\frac{1}{2}\left[\frac{ q_y^2\cos^2\tilde\theta}{l_-^2/l^4}+l^2_-(q_x\cos\tilde\theta+q_z\sin\tilde\theta)^2\right]} \nonumber\\
&\times L^2_n\left[\frac{1}{2}\left(\frac{ q_y^2\cos^2\tilde\theta}{l_-^2/l^4}+l^2_-(q_x\cos\tilde\theta+q_z\sin\tilde\theta)^2\right)\right],\label{eq_inactani}
\end{align}
where $l_+$ and $l_-$ are two typical lengths associated with the eigenvalues of the non-interacting Hamiltonian. They correspond to the characteristic length $\sqrt{\hbar/m\Omega}$ of the confining potential and the original magnetic length when the magnetic field is not tilted, respectively. The angle $\tilde\theta$ is a function of the tilt angle $\theta$ and
vanishes in the absence of tilting 
\begin{equation}
\tan 2\tilde\theta=\frac{\tan 2\theta}{1-\tan^2\theta-\frac{\Omega^2}{\omega_c^2}}.
\end{equation}

The interaction potential Eq.~\eqref{eq_inactani} is rather complicated. As we are pursuing a perturbative result, we can focus on the situation when the anisotropy is small. As the effective anisotropic interaction comes from the finite width in the $z$-direction and the tilt, the two most natural limits are the thin-sample limit and the small-tilt limit. The effective 2D interactions for the LLL in these limiting situations are listed below \cite{PhysRevB.96.195140}:
\begin{enumerate}
    \item Thin-sample limit: in this case, $\Omega\gg \omega_c$, and the effective interaction in the LLL takes the form:
    \begin{align}
        V_{\textrm{eff}}\simeq& 2\pi\frac{e^{-q^2l^2/2}}{q}\left(1-\sqrt{\frac{2\omega_c}{\pi\Omega}}ql\right.\nonumber\\
        &\left.+\frac{q^2-2q_x^2\tan^2\theta}{2(\Omega/\omega_c)}l^2\right),
    \end{align}
    where $\omega_c$ is the cyclotron frequency.
    \item Small-tilt limit: in this case, $B_x\ll B_z$, and the effective interaction in the LLL reads
    \begin{align}
         V_{\textrm{eff}}\simeq& 2\pi\frac{e^{-q^2l^2/2}}{q}\left\{A[z]+\frac{q^2_xl^2\tan^2\theta}{(\Omega/\omega_c)}\right.\nonumber\\
         &\left.\times\left[\sqrt{\frac{2\omega_c}{\pi\Omega}}ql-A[z]\left(1+\frac{q^2l^2\omega_c}{\Omega}\right)\right]\right\},
    \end{align}
    with $z=ql/\sqrt{2(\Omega/\omega_c)}$ and $A[z]=\exp(z^2)(1-\textrm{Erf}[z])$, in terms of the error function $\textrm{Erf}[z]$.
\end{enumerate}

Comparing the interaction with the expansion Eq. \eqref{eq_in_expan}, we find in these two limits that only $V_0$ and $V_2$ defined in Eq. \eqref{eq_in_expan} are non-vanishing. For the thin-sample limit, the expansion is
\begin{align}
    V_0(q)&=2\pi e^{-q^2l^2/2}\left(1-\sqrt{\frac{2\omega_c}{\pi\Omega}}ql+\frac{q^2l^2(1-\tan^2\theta)}{2(\Omega/\omega_c)}\right),\\
    V_2(q)&=-2\pi e^{-q^2l^2/2}\frac{q^2l^2\tan^2\theta}{2(\Omega/\omega_c)},
\end{align}
while for the small-tilt limit, we have
\begin{align}
    V_0(q)=&2\pi e^{-q^2l^2/2}\left\{A[z]+\frac{q^2l^2\tan^2\theta}{2(\Omega/\omega_c)}\right.\nonumber\\
    &\times\left.\left[\sqrt{\frac{2\omega_c}{\pi\Omega}}ql-A[z]\left(1+\frac{q^2l^2\omega_c}{\Omega}\right)\right]\right\},\\
    V_2(q)=&2\pi e^{-q^2l^2/2}\left[\sqrt{\frac{2\omega_c}{\pi\Omega}}ql-A[z]\left(1+\frac{q^2l^2\omega_c}{\Omega}\right)\right]\nonumber\\
    &\times\frac{q^2l^2\tan^2\theta}{2(\Omega/\omega_c)}.
\end{align}

\section{Neutral collective excitations under a tilted magnetic field}\label{sc_anicoll}

In this section, we present how to compute perturbatively the collective excitations of composite fermions under a tilted magnetic field. One may attempt to insert directly the anisotropic interaction into Eq. \eqref{eq_ve_iso}. However, this shifts the unphysical mode from zero to higher energies, and we show here how we can perturbatively perform the calculation while protecting the zero mode.

\subsection{Hartree-Fock Hamiltonian for CFs}\label{sc_HFH}

First we take a look at the CF Hamiltonian at the HF level in the presence of anisotropy. We show that the $V_{n\ne 0}$ components introduce a CF LL mixing term in the HF Hamiltonian.

The HF Hamiltonian is obtained by contracting the ladder operators into pairs, according to Eq. \eqref{eq_hf_repl}, in the interaction. The resulting Hamiltonian is
\begin{align}
    H_{CF}=&\frac{1}{2}\sum_{1234}\int \frac{d^2q}{(2\pi)^2}V(\mathbf q)\rho(\mathbf q)_{12}\rho(-\mathbf q)_{34}d^\dagger_1 d_2d^\dagger_3d_4\nonumber\\
    &\rightarrow \sum_{n}\epsilon_n d^\dagger_nd_n+\sum_{m\ne n}\left(M_{m,n}d^\dagger_md_n+\textrm{c.c.}\right),
\end{align}
where the indices $m,n$ now only label the CF LL. The implicit intra-LL indices in the right-hand side of the above formula are diagonal and thus omitted. The HF energy $\epsilon_n$ appears naturally in the HF Hamiltonian and is accompanied with CF-LL mixing amplitudes $M_{m,n}$:
\begin{align}
    M_{m,n}&=\int \frac{d^2q}{4\pi^2} \frac{V(\mathbf q)}{2}\bigg[\delta_{m,n}
    -2\sum_{l=0}N_F(l)
    \rho_{ln}(\mathbf q)\rho_{ml}(-\mathbf q)\big]\nonumber\\
    &=\frac{1}{2}\int \frac{qdq}{2\pi} \left[V_0(q)\delta_{m,n}-\sum_{l=0}N_F(l)(1+\delta_{m,n})\right.\nonumber\\
    &\quad\times \left.\rho_{ln}(q)\rho_{ml}(-q)V_{|m-n|}(q)\right],\label{eq_hmani}\\
    \epsilon_n&=M_{n,n},
\end{align}

From the above expression, we can see that for an isotropic interaction, there is no mixing between different CF LLs. In this case, the CF indeed behaves like free charged particles moving in an effective magnetic field and the CF-LL structure is self-consistent. The charge gap, the energy to create a well-separated quasiparticle/hole pair, is to add up the energy to create a quasihole in the topmost filled CF LL and a quasiparticle in the lowest empty one (using the preferred density), as we have discussed above.

When we are confronted with anisotropy, the mixing amplitude changes the problem significantly. In principle the ground state can no longer be viewed as $p$ filled CF LLs. However, in our previous work \cite{PhysRevB.98.205150}, we showed that in the calculation of the activation gap, we can eliminate such mixing near the CF Fermi level by choosing an optimal geometry for the CF cyclotron orbitals. Then we can still compute the activation gap as if the mixing is turned off. But for the calculation of the collective excitations, we have to consider a large number of CF LLs so that the mixing cannot be eliminated easily. In the next section, we show how a perturbative method can be constructed to extract information about the collective excitation.

\subsection{The equation of motion for an anisotropic interaction}

In perturbation theory, we still evaluate the density correlation function in terms of CF-operator averages over the reference state $|\mathbf p\rangle$, which consists of
$p$ filled CF LLs, even if the latter is not a perfect candidate for the true ground state because of the CF LL mixing described above. If one computes the equation of motion for the one-body Green's function, the empty CF LL will be occupied by CFs from the filled CF LLs. But as we are dealing within the perturbative regime, it may still be reasonable to use it to evaluate the density correlation. Here we compute the Fourier transformation $\int dt\exp(i\omega t)\langle\mathbf p|\mathcal T \rho(t,\mathbf r)\rho(0,0)|\mathbf p\rangle$. To see its usefulness, we need the decomposition into the eigenstates of the Hamiltonian:
\begin{align}
    \Pi(\omega,\mathbf q)=&\sum_{\alpha,\beta}\frac{\langle \mathbf p|\alpha\rangle\langle \alpha|\rho(\mathbf q)|\beta\rangle\langle \beta|\rho(-\mathbf q)|\mathbf p \rangle}{\omega-(E_\beta-E_\alpha)+i0^+}\nonumber\\
    &-\frac{\langle \mathbf p|\rho(-\mathbf q)|\beta\rangle\langle \beta|\rho(\mathbf q)|\alpha\rangle\langle \alpha|\mathbf p\rangle}{\omega-(E_\alpha-E_\beta)-i0^+},
\end{align}
where $|\alpha\rangle$ and $|\beta\rangle$ represent eigenstates of the system. From this decomposition, one can see that, if the ground state has a significant overlap with that of $p$ filled CF LLs, the poles of this density correlation function  still provide information about the neutral excitations. Using this state is therefore valid in the perturbative regime.

The evaluation of the equation of motion is done as before, by computing the commutator between the Hamiltonian and the magneto-exciton operator and approximating a pair of creation and annihilation operators with their expectation value. However, this time the expectation value is not conserved self-consistently due to the CF-LL mixing $\langle d^\dagger(t)d(t)\rangle\ne \langle d^\dagger(0)d(0)\rangle$. Empty CF LLs are be populated and filled CF LLs will loose CFs during the time evolution. From a perturbative point of view, if we treat the CF LLs as a true free-particle structure and not  as an emergent one, this perturbation should be oscillating in time with a small amplitude \cite{landau1962quantum}. As a first step towards the anisotropic perturbation, we thus still put the time-dependent occupation expectation to be its value at $t=0$
\begin{equation}
    \langle d^\dagger_{n_1,X_1}(t)d_{n_2,X_2}(t)\rangle\simeq \delta_{X_1X_2}\delta_{n_1n_2}N_F(n_1).
\end{equation}
The calculation for the equation of motion is now almost identical to the isotropic case, except that the mixing matrix $M_{m,n}$ appears in addition to the original CF LL energy difference $\epsilon_{n_1}-\epsilon_{n_2}$, scattering one leg of $O_{n_1n_2}$.  One finds that the equation now takes the form:
\begin{align}
      \omega \langle O_{n_1n_2}O_{n^{\prime\prime}_2n^{\prime\prime}_1}\rangle=&\sum_{n'_1,n'_2}\left(M_{n'_1,n_1}\delta_{n_2,n'_2}-M_{n_2,n'_2}\delta_{n_1,n'_1}\right) \nonumber\\
      &\times\langle O_{n'_1n'_2}O_{n^{\prime\prime}_2n^{\prime\prime}_1}\rangle+[N_F(n_2)-N_F(n_1)]\nonumber\\
    &\times\sum_{n'_1,n'_2}\left(V^{(2)}_{n_2,n'_1,n'_2,n_1}-V^{(1)}_{n_2,n'_1,n_1,n'_2}\right)\nonumber\\
    &\times\langle O_{n'_1n'_2}O_{n^{\prime\prime}_2n^{\prime\prime}_1}\rangle-iN_{\phi^\ast}\delta_{n_1,n^{\prime\prime}_1}\delta_{n_2,n^{\prime\prime}_2}\nonumber\\
    &\times[N_F(n_1)-N_F(n_2)].\label{eq_ve_ani}
\end{align}
This equation is different from Eq.~\eqref{eq_ve_iso} by the appearance of the amplitudes $M_{m,n}$ with $m\ne n$. With their inclusion, one can verify that the above matrix equation admits a left zero-energy eigenvector, still given by $\chi_{n_1n_2}(\mathbf q)$. The details can be found in Appendix \ref{ap_zmode}. This merit enables us treat Eq.~\eqref{eq_ve_iso} as a good perturbation equation.

\begin{figure*}
        \centering
        \begin{minipage}[t]{0.47\linewidth}
        \includegraphics[width=\linewidth]{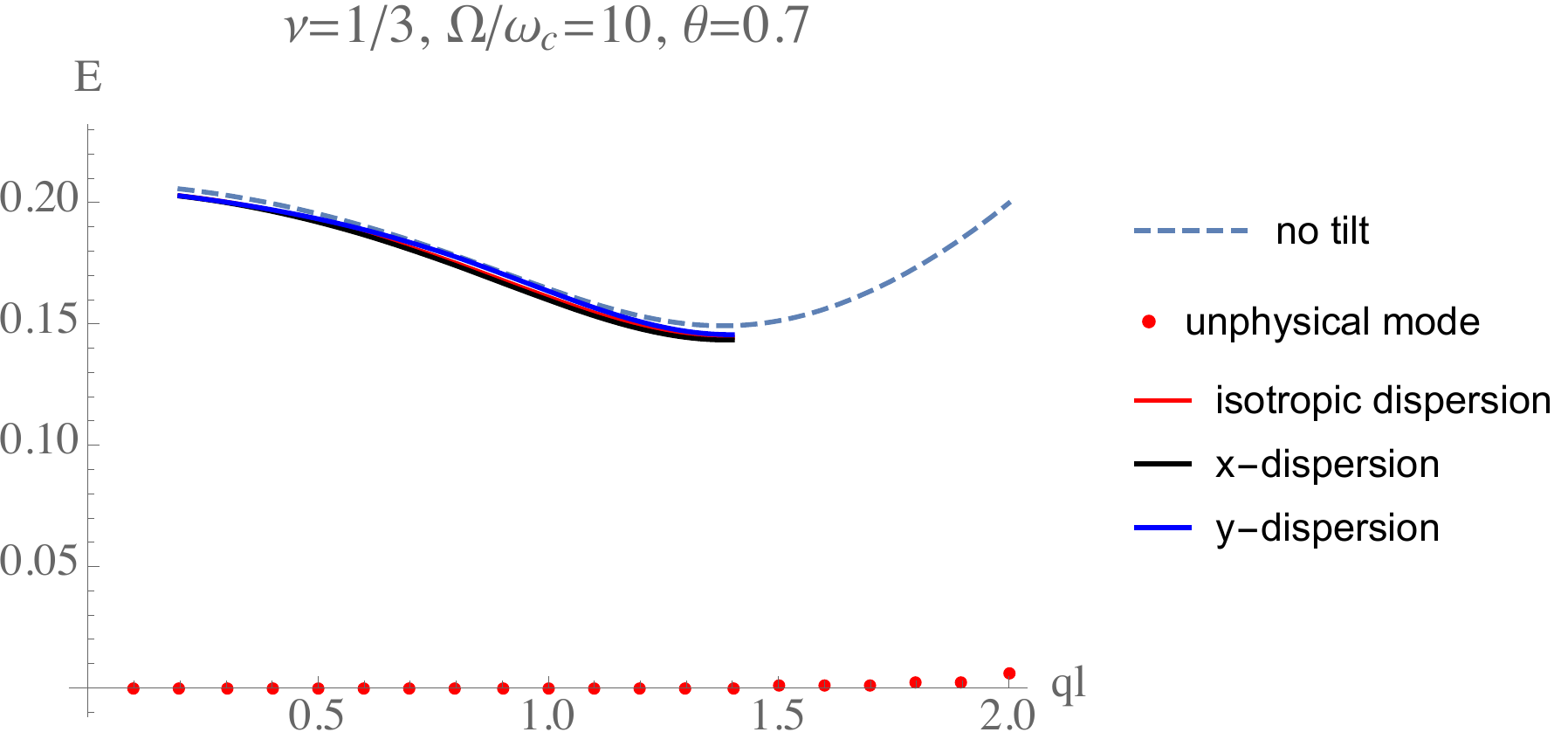}
        \subcaption{}\label{fig_dis3_s}
        \end{minipage}
        \begin{minipage}[t]{0.47\linewidth}
        \includegraphics[width=\linewidth]{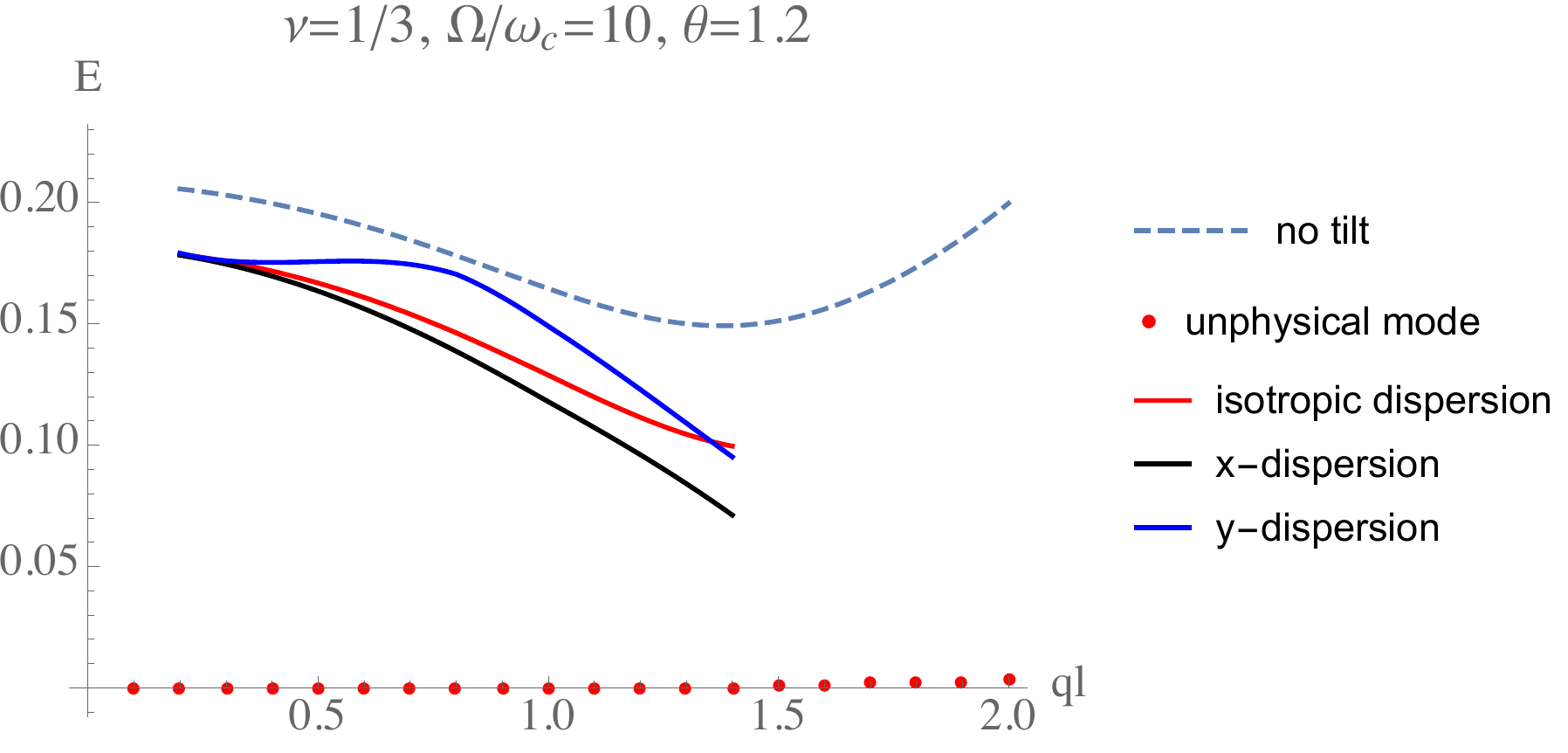}
        \subcaption{}\label{fig_dis3}
        \end{minipage}

        \begin{minipage}[t]{0.47\linewidth}
         \includegraphics[width=\linewidth]{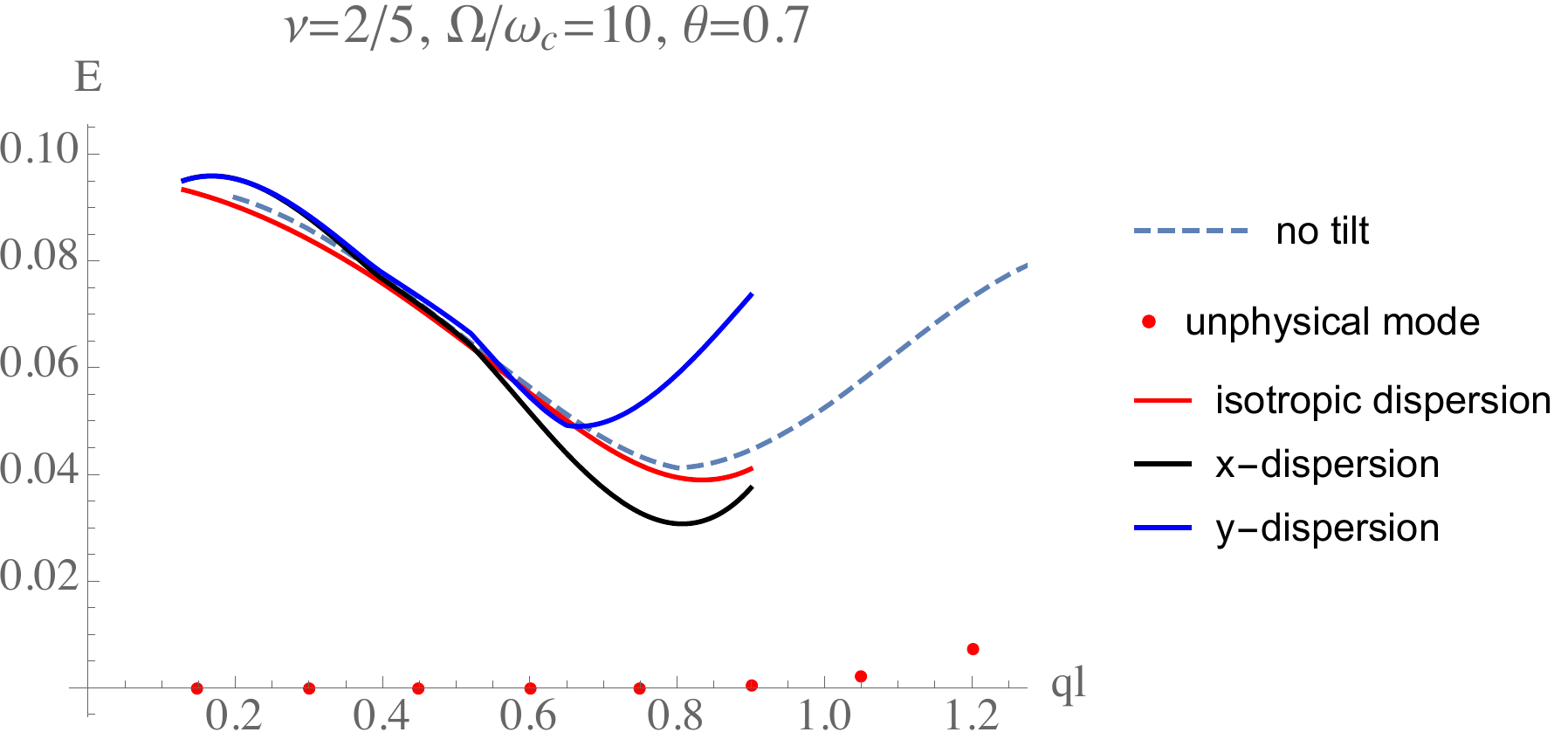}
        \subcaption{}\label{fig_mexp_com}
        \end{minipage}
        \begin{minipage}[t]{0.47\linewidth}
         \includegraphics[width=\linewidth]{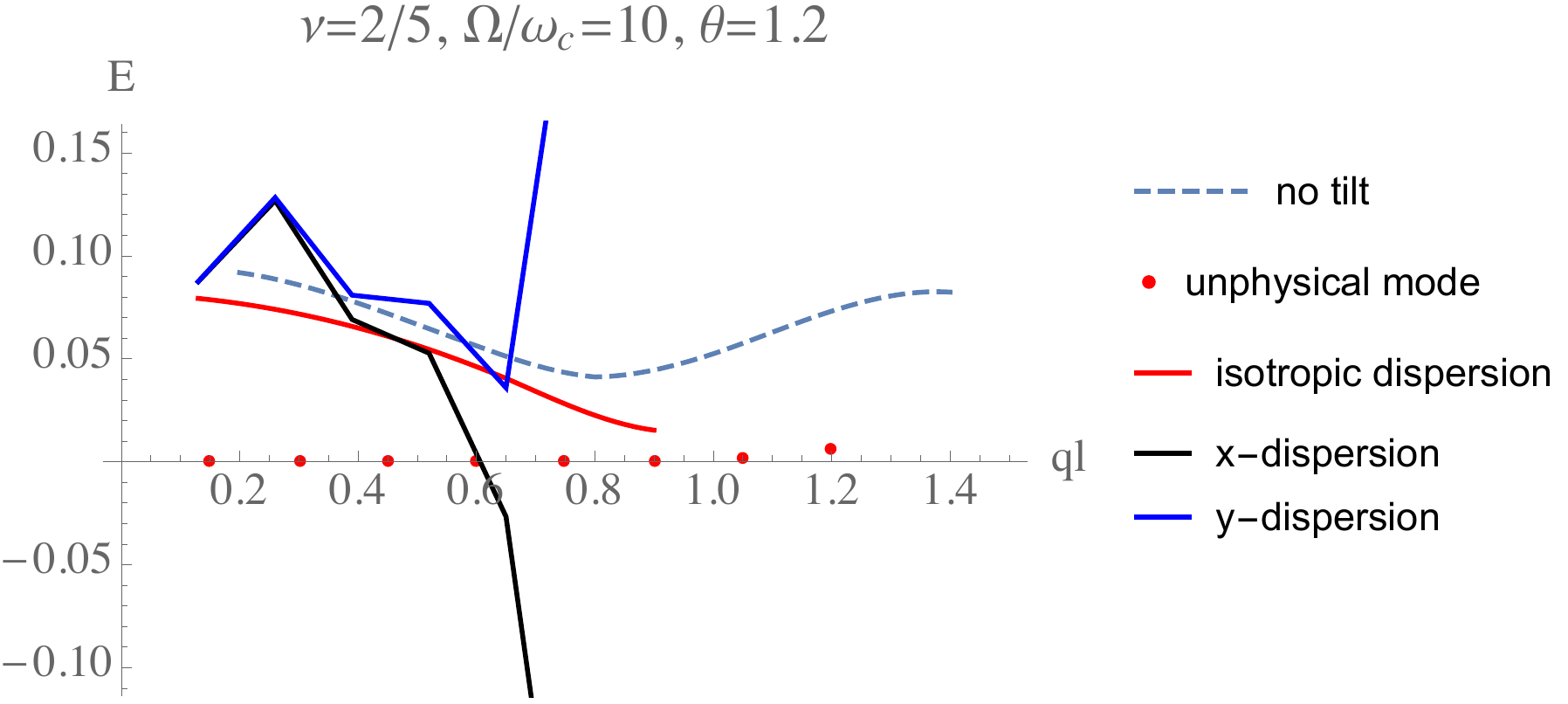}
        \subcaption{}\label{fig_mexp_com_l}
        \end{minipage}
        \caption{Dispersion of the collective excitations at different fillings and $\theta$, for (a) $\theta=0.7$, $\nu=1/3$; (b) $\theta=1.2$, $\nu=1/3$; (c) $\theta=0.7$, $\nu=2/5$; (d) $\theta=1.2$, $\nu=2/5$.
        Due to computational difficulty, we only calculate the anisotropic dispersion up to the first minimum corresponding to the magneto-roton gap: $ql\simeq 1.4$ for $\nu=1/3$ and $ql\simeq 0.8$ for $\nu=2/5$. The isotropic dispersion (red lines) is obtained by considering merely the influence of the tilt on the isotropic component $V_0$ of the interaction potential, while the dispersion in the $x$ (black lines) and $y$ (blue liens) directions take into account the $V_{n\ne 0}$ components. }
\end{figure*}

As before we group the above matrix equation into a block form according to the types of excitons classified in Sec. \ref{sc_TDHF}. Due to the matrix $M_{m,n}$, there is a mixing between type-\ref{tp_phy} excitons and type-\ref{tp_un}, type-\ref{tp_uns} excitons. The block matrix equation is expressed as:
\begin{equation}
    \left(\begin{array}{ccc}
        \omega-A & -B  \\
        -C & \omega-D
    \end{array}\right) \left(\begin{array}{ccc}
        a & b  \\
        c & d
    \end{array}\right)=
     \left(\begin{array}{ccc}
        \tilde I & 0 \\
        0 & 0
    \end{array}\right),
\end{equation}
The vectors $b$ and $d$ satisfy two homogeneous matrix equations, $(\omega-A)b-Bd=0$ and $Cb-(\omega-D)d=0$. In general they have to be zero as there are usually no common solutions to both equations. For the same reasons that we invoked before, this is also easily understood because type-\ref{tp_un} and type-\ref{tp_uns} exciton operators annihilate the reference state. The block $c$ is not an independent variable and is given by $c=-(D-\omega)^{-1}Ca$. The above matrix equation~\eqref{eq_ve_ani} then transforms to
\begin{equation}
    \left[(\omega-A)-B(\omega-D)^{-1}C\right]a=\tilde I.\label{eq_ani_veeq}
\end{equation}
Compared to the isotropic case, we have an $\omega$-dependent perturbation. The poles of the correlation function are given by the condition $\det[(\omega-A)-B(\omega-D)^{-1}C]=0$.

Now we solve this equation perturbatively. The influence of the tilt on $V_0$ is taken non-perturbatively while $V_{n\ne 0}$ are the perturbations. The matrix $C$ is vanishing when there is no $V_{n\ne 0}$ component. So to leading order of perturbation expansion, we need to keep $A^{(0)}$, $A^{(1)}$, $B^{(0)}$, $C^{(1)}$ and  $D^{(0)}$, where $A^{(0)}$,  $B^{(0)}$, and $D^{(0)}$ are computed from $V_0$ and $A^{(1)}$ and $C^{(1)}$ are the leading order perturbation induced by $V_{n\ne 0}$.  Equation~\eqref{eq_ani_veeq} does not take the form of an eigenvalue equation. But to compute the leading order perturbation, we can simply replace $\omega$ in $(\omega-D)^{-1}$ by $\omega^{(0)}$, the eigenvalue of the isotropic $A^{(0)}$ such that the equation again takes an eigenvalue form, which we denote as $(\omega-\mathcal H)a=\tilde I$, where
\begin{equation}
    \mathcal H=A^{(0)}+A^{(1)}-B^{(0)}(\omega^{(0)}-D^{(0)})^{-1}C^{(1)}.
\end{equation}

With these equations we perform a standard perturbation calculation for the eigenvalues. Notice that here the matrix $\mathcal H_{\mu,\nu}$ is not Hermitian. We have to distinguish left and right eigenvectors. According to non-degenerate perturbation theory, the correction to the eigenvalue to leading order can in this case be given by the expectation value of the perturbation matrix. We thus need the eigenstates of the first positive non-zero eigenvalue, denoted as $|\Psi^R_1\rangle$ and $\langle\Psi^L_1|$. The matrix $\mathcal H$ is decomposed into $\mathcal H=A^{(0)}+\mathcal H'$, with $\mathcal H'$ the difference between $A^{(0)}$ and $\mathcal H$. Perturbation theory for a non-degenerate eigenstate tells us that the correction to the energy is given by:
\begin{equation}
    \delta E(\mathbf p)=\frac{\langle\Psi^L_1|\mathcal H'|\Psi^R_1\rangle}{\langle\Psi^L_1|\Psi^R_1\rangle}.
\end{equation}

In Figs.~\ref{fig_dis3_s} and \ref{fig_dis3} we show the results for $\nu=1/3$ in the thin-sample limit. We plot both the dispersion computed from the $V_0$ component in the presence of tilt (denoted as isotropic dispersion) and the dispersion along the $x$- or the $y$-direction by adding the perturbation from the $V_{n\ne 0}$ components. The dispersion in $q_y$-direction (blue lines) is increased while that in the $q_x$-direction (black lines) is lowered. This means that the exciton pair separated by $y\sim q_xl^2$ is more likely to condensate. This may be interpreted in terms of the cyclotron motion : remember that the inplane component of the magnetic field is in oriented along the $x$-direction, such that the (tilted) wavefunctions are compressed in this direction when projected back to the $xy$-plane. Furthermore, due to the tilt of the wavefunctions, the overlap between adjecent particles is decreased in the $x$-direction while that in the $y$-direction remains strong.  The quasiparticle/hole pair the components of which are separated in the $y$-direction can therefore benefit better from the exchange-interaction-induced attraction in the region of overlapping wavefunctions. In addition to providing an anisotropic perturbation, the tilt also lowers the overall magneto-roton gap by modifying the isotropic interaction $V_0$ for large tilt angle, as shown by the red lines in Figs.~\ref{fig_dis3_s} and \ref{fig_dis3}. In this perturbation calculation, the collective excitation for $\nu=1/3$ exhibits a very strong robustness. The perturbation is one order of magnitude smaller than the activation gap for even very large tilt angle. This is qualitatively consistent with exact-diagonalization calculations in Ref.~\cite{Yang_2018}. There the $\nu=1/3$ spectrum gap is not closed until $B_x/B_z=5$, which is beyond the regime where the perturbation theory can be applied.

In contrast, the neutral collective excitation for $\nu=2/5$ is much vulnerable. The results are summarized in Fig. \ref{fig_mexp_com}. At around $\theta=0.7$ and $\Omega/\omega_c=10$, the anisotropic perturbative contribution is already of the same magnitude as the magneto-roton gap. In principle, the perturbation theory should not be valid from this point on, but this still provides useful information for the behaviour of the system. Furthermore our calculation for larger angle $\theta=1.2$ indicates that the magneto-roton gap can be closed by tilting the magnetic field. In Fig.~\ref{fig_mexp_com_l}, the anisotropic perturbation starts to diverge. The results become highly irregular and the system is not stable under such anisotropy. Combined with our previous results on the remarkable robustness of the activation gap \cite{PhysRevB.98.205150}, we find indications for a different, possibly nematic, phase that has the same type of topological order as the $\nu=2/5$ CF state but that possesses different geometric properties.

\section{Conclusions}

In summary, we have generalized the TDHF approximation of CFs perturbatively to take into account anisotropic interactions. The approach allows us to compute the spectrum of neutral collective excitations when the magnetic field is tilted from the direction perpendicular to the 2D quantum well. The $\nu=1/3$ state tends to exhibit a robustness in the perturbative regime while the $\nu=2/5$ state behaves very vulnerably. In our calculation we show that the perturbation for the latter exhibits divergences around $\theta\sim1.2$. The magneto-roton gap is very likely to close for such a large tilt angle. In common uniform liquid systems, this kind of finite wave-vector gap closing leads to charge density wave states. In the LLL, the charge density wave is usually not stable against quantum fluctuations. A very promising candidate is the nematic phase with an intrinsic director order \cite{PhysRevB.88.125137,PhysRevX.4.041050}. This state is suggested as the candidate after the collective mode is softened \cite{PhysRevX.7.041032,PhysRevB.98.155140}. Our results signify that there may be a nematic phase transition by increasing the tilt angle.

We also study the large wave-vector behaviour of the collective spectrum in the Hamiltonian theory. We compare the spectrum obtained by the preferred density and electron density. The former gives the correct charge gap but fails to reproduce the correct dispersion. This can be explained as follows. The small momentum behaviour is related to short-distance magneto-exciton pairs of CFs. It is the electron density instead of the effective CF density that is seen at such short distances. This needs to be contrasted to the large wave-vector limit where the collective excitations can be viewed as a pair that consists of a CF that is well separated from the CF hole. At these large wave vectors, the short-distance fine structure of the CF is no longer relevant, and the CF can be effectively viewed as a pointlike particle with an effective charge $e(1-c^2)$, as it is described in the preferred CF density. The preferred density is therefore the appropriate quantity to describe CFs in this limit.

In the small wave-vector limit, we need to appeal to the electronic density along with the physical constraint, and the Hamiltonian theory provides an additional, albeit unphysical, zero mode in the collective-excitation spectrum. The presence of this zero mode can be used as a benchmark in practical calculations, but to maintain it at zero energy requires a rapidly increasing number of CF LLs in the study of the spectrum at large wave vectors. As consequence, the dispersion computed from the electron density with a fixed cutoff does not converge rapidly to the CF-LL spacing calculated with the help of
the preferred density, which is more reliable in the large wave-vector limit.

Based on what we have obtained, one may wonder how we can go beyond the perturbation study in this paper. There are still several questions and challenges. The first is to obtain a better understanding of the CF-LL mixing in terms of the original CF wavefunction approach. It turns out that the Laughlin-Jastrow factors in anisotropic trial FQH states \cite{PhysRevB.85.115308,PhysRevB.93.075121} no longer take a vortex form. It may be interesting to investigate how this is related to the CF-LL mixing found in the Hamiltonian theory. Second, in order to go beyond perturbation theory, we need to solve numerically, by iteration, a self-consistent HF reference state so that there is no level mixing. This state should have a very different CF occupation pattern $\langle d^\dagger d\rangle$ from that of the reference state with $p$ filled CF LLs. It is not clear whether the numerical computation can converge sufficiently fast in practice. On the other hand, there is more and more evidence \cite{son2019chiral} that in anisotropic quantum Hall systems the Dirac CFs \cite{PhysRevX.5.031027} capture all essential geometric information naturally. It would be interesting to investigate the self-consistently calculated reference state in terms of the Dirac CF picture.

\begin{acknowledgments}
We thank Nicolas Regnault, Zlatko Papi\'c, Ajit Balram, Ady Stern and Steve Simon for helpful discussions.
\end{acknowledgments}

\appendix

\section{The zero mode in anisotropic case}\label{ap_zmode}

In this part, we show that a zero mode exists for Eq.~\eqref{eq_ve_ani}, which is exactly the pseudovortex density imposing the physical constraint. The symbol $\rho$ used in this appendix always refers to the electron density $\rho_e$.

It is easier to work out the zero left-eigenvector of Eq.~\eqref{eq_ve_ani}. This can be illustrated as a consequence that the HF average of a vanishing operator should also vanish. If we insert $\sum\chi_{n_1n_2} O_{n_1n_2}$ into the commutator with $\sum\rho_{n_1n_2}O_{n_1n_2}$, it gives zero and this should persist after performing the Wick contraction. Before performing the calculation, we need an identiy:
\begin{align}
    \sum_n \chi_{n_1n}(\mathbf q_1)\rho_{nn_2}(\mathbf q_2)= &e^{-i\mathbf q_1\times\mathbf q_2l^{\ast2}}\sum_n\rho_{n_1n}(\mathbf q_2)\nonumber\\
    &\times\chi_{nn_2}(\mathbf q_1).\label{eq_exchrh}
\end{align}
This can be easily verified by writing out explicitly the operator forms of $\chi$ and $\rho$.

Now we verify the zero mode. For the anisotropic case, the excitation matrix for excitons takes the form:
\begin{align}
    &M_{n'_1,n_1}\delta_{n_2,n'_2}-M_{n_2,n'_2}\delta_{n_1,n'_1}+[N_F(n_2)-N_F(n_1)]\nonumber\\
    &\left(V^{(2)}_{n_2,n'_1,n'_2,n_1}-V^{(1)}_{n_2,n'_1,n_1,n'_2}\right),\label{eq_animxfm}
\end{align}
which we multiply from the left with $\chi_{n_1n_2}$. The explicit forms of $M_{m,n},V^{(1)}$ and $V^{(2)}$ are given by Eqs.~\eqref{eq_hmani},~\eqref{eq_ve_vexc1}, and~\eqref{eq_ve_vdir1}. The CF-LL mixing terms give the contribution:
\begin{equation}
    \sum_{n}\chi_{nn'_2}M_{n'_1,n}-\chi_{n'_1n}M_{n,n'_2}.\label{eq_hpctb}
\end{equation}
The product between the direct interaction $V^{(2)}$ and $\chi_{n_1n_2}$ gives a vanishing contribution:
\begin{align}
    &\sum_{n_1,n_2}\left[N_F(n_2)-N_F(n_1)\right]\chi_{n_1n_2}V^{(2)}_{n_2n'_1n'_2n_1}\nonumber\\
    =&\sum_{n_1,n_2}\frac{1}{2\pi l^{\ast2}}\left[N_F(n_2)-N_F(n_1)\right]\chi_{n_1n_2}(\mathbf q)\nonumber\\
    &\times V_{\textrm{eff}}(\mathbf q)\rho_{n'_1n'_2}(\mathbf q)\rho_{n_2n_1}(-\mathbf q)\nonumber\\
    =&\sum_{n_1,n_2}\frac{1}{2\pi l^{\ast2}} V_{\textrm{eff}}(\mathbf q))\rho_{n'_1n'_2}(\mathbf q)\big [N_F(n_2)\chi_{n_1n_2}(\mathbf q)\nonumber\\
    &\times \rho_{n_2n_1}(-\mathbf q)-N_F(n_1)\chi_{n_2n_1}(\mathbf q)\rho_{n_1n_2}(-\mathbf q)\big]\nonumber\\
    =&0,
\end{align}
where in the second step we make use of Eq.~\eqref{eq_exchrh}. For the exchange interaction, still making use of Eq.~\eqref{eq_exchrh}, the term times $N_F(n_2)$ becomes
\begin{align}
    &\sum_{n_1,n_2}-N_F(n_2)\chi_{n_1n_2}(\mathbf q)\int \frac{d^2\mathbf k}{(2\pi)^2}V_{\textrm{eff}}(\mathbf k)e^{i\mathbf k\times \mathbf ql^{\ast2}}\nonumber\\
     &\times\rho_{n'_1n_1}(\mathbf k)\rho_{n_2n'_2}(-\mathbf k)\nonumber\\
     =&-\sum_{n_1,n_2}\chi_{n'_1 n_1}(\mathbf q)\int \frac{d^2\mathbf k}{(2\pi)^2}V_{\textrm{eff}}(\mathbf k)N_F(n_2)\rho_{n_1n_2}(\mathbf k)\nonumber\\
     &\times\rho_{n_2n'_2}(-\mathbf k).
\end{align}
The integral is exactly the CF-LL dependent part in $M_{m,n}$ [Eq.\eqref{eq_hmani}]. The above expression is equal to $\chi_{n'_1n}M_{n,n'_2}$ up to a CF-LL independent term. The exchange interaction times $N_F(n_1)$ result in a similar term for $-\chi_{nn'_2}M_{n'_1,n}$. As the CF-LL independent parts, $(1/4\pi)\int qdqV_0(q)\delta_{m,n}$ in  $M_{m,n}$, cancel in the difference Eq.~\eqref{eq_hpctb}, the exchange interaction just gives:
\begin{equation}
    \sum_{n}\chi_{n'_1n}M_{n,n'_2}-\chi_{nn'_2}M_{n'_1,n}.
\end{equation}
So it cancels the contribution from the CF-LL mixing matrices Eq.~\eqref{eq_hpctb}. The pseudovortex density $\chi_{n_1n_2}$ is a left zero eigenvector of Eq.~\eqref{eq_animxfm}.

\nocite{*}

\bibliography{apssamp}

\end{document}